\documentclass[runningheads]{llncs}
\usepackage[T1]{fontenc}
\usepackage{graphicx}
\usepackage{bbding}
\usepackage{amsmath}

\DeclareMathOperator*{\argmin}{arg\,min}

\usepackage{subfigure}
\begin{document}
\title{Low-dose CT reconstruction by self-supervised learning in the projection domain}


%
%
\author{Long Zhou\inst{1*} 
	\and
	Xiaozhuang Wang\inst{2*}
	\and
	Min Hou\inst{1}
	\and
	Ping Li\inst{1,3}
	\and
	Chunlong Fu\inst{4}
	\and
	Yanjun Ren\inst{2}
	\and
	Tingting Shao\inst{1}
	\and
	Xi Hu\inst{1}
	\and
	Jihong Sun\inst{1}
	\and
	Hongwei Ye\inst{2}\textsuperscript{\Envelope}
}
\authorrunning{L. Zhou et al.}
%
\institute{Sir Run Run Shaw Hospital, Zhejiang University School of Medicine, Hangzhou, China \\ \email{zhoulong21@zju.edu.cn}\and
	ZheJiang Minfound Intelligent Healthcare Technology Co., Ltd., West Wenyi road, Hangzhou, China \\ \email{Hongwei.Ye@minfound.com} \and Zhejiang Chinese Medical University Affiliated Jiaxing TCM Hospital, China \and Dongyang Hospital of Wenzhou Medical University, China 
}
\maketitle              
\begin{abstract}

In the intention of minimizing excessive X-ray radiation administration to patients, low-dose computed tomography (LDCT) has become a distinct trend in radiology. However, while lowering the radiation dose reduces the risk to the patient, it also increases noise and artifacts, compromising image quality and clinical diagnosis. In most supervised learning methods, paired CT images are required, but such images are unlikely to be available in the clinic. We present a self-supervised learning model (Noise2Projection) that fully exploits the raw projection images to reduce noise and improve the quality of reconstructed LDCT images. Unlike existing self-supervised algorithms, the proposed method only requires noisy CT projection images and reduces noise by exploiting the correlation between nearby projection images. We trained and tested the model using clinical data and the quantitative and qualitative results suggest that our model can effectively reduce LDCT image noise while also drastically removing artifacts in LDCT images.

\end{abstract}
\section{Introduction}
Computed tomography (CT) scans are commonly used in hospitals all over the world. However, the presence of x-rays in the scans exposes persons to ionising radiation, which can cause genetic damage and cancer induction. Low-dose CT (LDCT) imaging employs less radiation, reducing damage but increasing noise and artifacts in reconstructed images. These contaminated CT images with artifacts, which may adversely affect their clinical usefulness scan \cite{2009Computed,1987Noise}.

Low-flux acquisition is a common approach for reducing X-ray dose in a single exposure by modifying the X-ray tube current or exposure period \cite{2004Strategies}. Low X-ray exposure will result in noisy projection, resulting in noisy CT images. Several LDCT reconstruction algorithms have been proposed in recent decades to obtain high-quality LDCT images. Sinogram filtering \cite{2009Projection,2005Sinogram,2012Image}, iterative reconstruction \cite{2012Iterative,2009Iterative}, and image post-processing \cite{dutta2013non,2013Image} are the three categories in which they fall.

With the renaissance of artificial intelligence in the past decade, various deep neural networks were proposed to denoise LDCT images, which became the main stream methods\cite{Hu2017Low,2018Low,20183D}. On the other hand, the majority of previous deep learning approaches for LDCT denoising depend on supervised learning with the normal-dose CT (NDCT) image as a ground truth of LDCT images \cite{Hu2017Low,2018Low,20183D}. In a clinical setting, acquiring such paired data is very challenging due to the higher X-ray dose from multiple scans, as well as the patient's breathing and motion, which can cause multiple scan to be misaligned. Because getting paired data is difficult, most image-domain methods build paired images by adding noise to the NDCT images. However such an approach can bias the predictions of a CNN because no synthetic noise can completely emulate the real one to portray all physical phenomena encountered in a CT machine.
 
Recent studies on self-supervised learning have demonstrated that denoising networks can be trained without the use of clean references \cite{2018Noise2Noise,2019Noise2Self,2019Noise2Void}. In these methods, the noiseless signal is deduced from the noisy image itself. A pioneering work known as Noise2Noise training \cite{2018Noise2Noise} has shown that training a denoising network from pairs of noisy images. This Noise2Noise method was applied for denoising X-ray projections and CT images and compared to a supervised model in \cite{gnudi2020denoising}. It gave acceptable results but the images were over-smoothed. Most often, only one noisy image of a particular scene is available. As a step further, self-supervised learning, which requires neither clean targets nor noisy image pairs in denoising tasks, has been proposed \cite{2019Noise2Self,2019Noise2Void,2019High,papkov2021noise2stack,zhang2021noise2context}, but they add complexity to the model, and training process making it less efficient. In recent work \cite{zainulina2021no}, a self-supervised learning models have been developed to do projection data noise reduction with a promising result. However, the method has only been evaluated on simulated data and has yet to be validated on clinical data.

Inspired by previous work \cite{papkov2021noise2stack,zhang2021noise2context,zainulina2021no} and our understanding of CT imaging. We propose to employ a self-supervised learning model,Noise2Projection, to do low-dose CT image reconstruction using raw projection data. We directly used clinical data to train and test the model, and the quantitative and qualitative results suggest that our model can effectively reducing LDCT image noise while also dramatically remove artifacts in LDCT images.

\section{Methodology}
\subsection{Noise2Projection}
\begin{figure*}[htb]
	\begin{center}
		\includegraphics[scale=0.4]{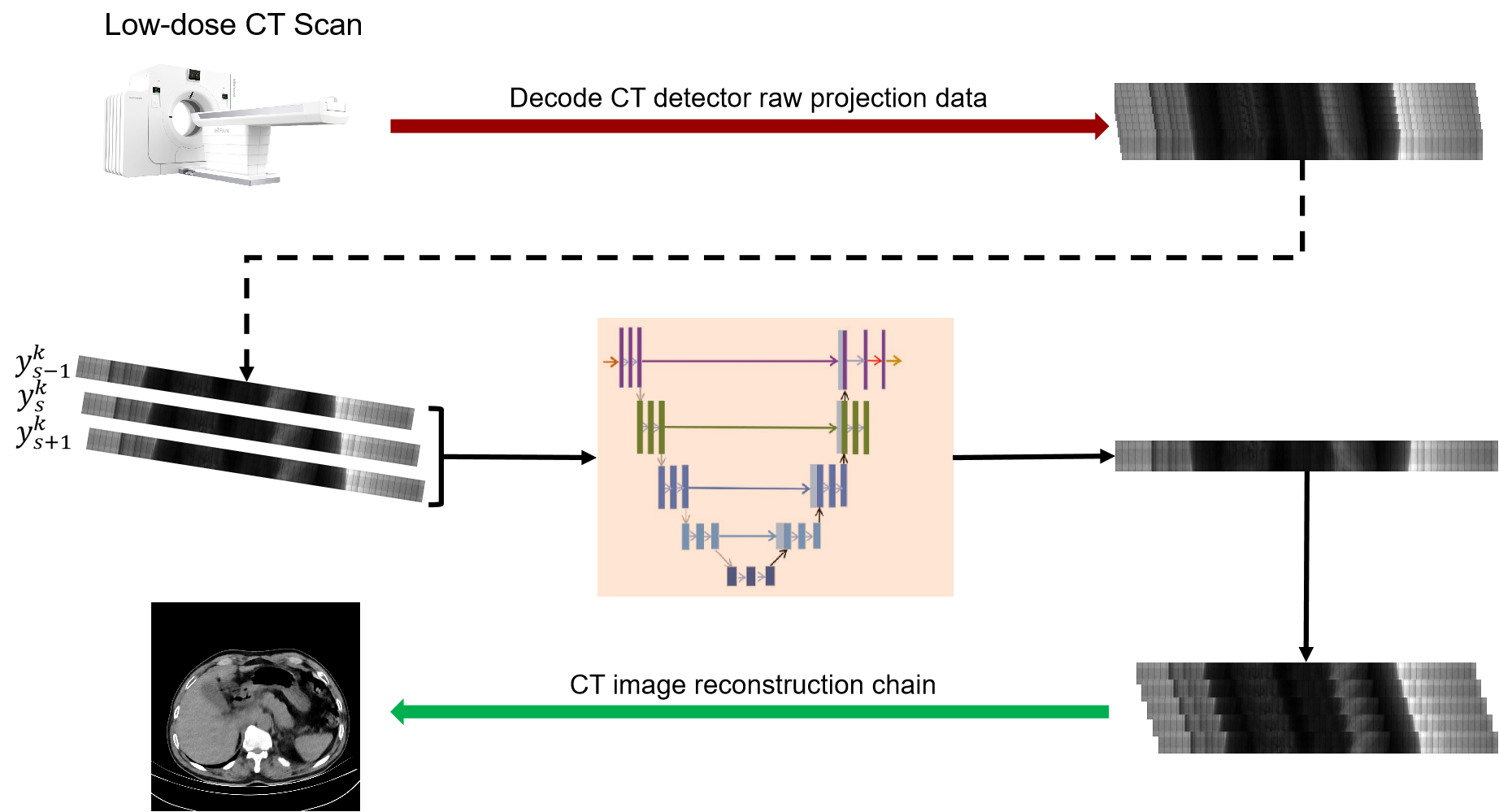}
		\caption{The flowchart of the proposed method.To begin, we get the raw data from the clinical LDCT scan and decode it into projection data. The noise derived projection data is then obtained by randomly cropping three neighbouring angle projection data ($y^k_{s-1}$,$y_{s}^k$ and $y_{s+1}^k$) and feeding it into our model. Finally, CT images were reconstructed from the denoised projection data.}
		\label{network}
	\end{center}
\end{figure*}

Inspire by Noise2Context\cite{zhang2021noise2context} and  Noise2Stack\cite{papkov2021noise2stack}, we composed above two methods and reconstructed our model, Noise2Projection. Because there are over 1000 projections in one rotation of our CT scan, the variations between neighbouring projection data are minimal. As a result, this model for projection data is being considered in order to reduce noise and increase the quality of reconstructed CT images. We use the projection data of three consecutive angles as the model's input and estimate the projection data of the intermediate angles, as indicated in Fig.\ref{network}.

For LDCT denoising in the projection domain, the basic mathematical model can be formulated as: $y = x + n$, $y$ is the noisy LDCT projection image, $x$ is the corresponding clear reference, $n$ denotes the noise. To obtain clean LDCT projection image from $y$, a neural network $F_{\theta}$ can be trained with object function as in Eq.\ref{eq2}:

\begin{equation}
	\theta = \argmin\limits_{\theta}\sum_{}^{}{\left|\left| F_{\theta}(y) - x\right|\right|_2^2}
	\label{eq2}
\end{equation}

In our model, we use the projection data of three adjacent angles ($y^k_{s-1}$,$y_{s}^k$ and $y_{s+1}^k$) as the input to the model and estimate the projection data of the intermediate angles. Since there is no corresponding clean projection images, we use a self-supervised learning method to train our model. The objective function of our network is Eq. \ref{eq1}. 

\begin{equation}
	\begin{aligned}
		\theta &= \argmin\limits_{\theta}\sum_{k=0}^{K}\sum_{s=1}^{S-1}{\left|\left| F_{\theta}(y_s^k) - y_{s-1}^k\right|\right|_2^2 + \left|\left| F_{\theta}(y_s^k) - y_{s+1}^k\right|\right|_2^2 + \left|\left| F_{\theta}(y_s^k) - y_{s}^k\right|\right|_2^2} \\
	\end{aligned}
	\label{eq1}
\end{equation}

Where $y_s^k$, $y_{s-1}^k$, and $y_{s+1}^k$ are three adjacent projection data of the $k$th patient, $s$ is the index of projection image. 

Similar to the \cite{zhang2021noise2context}, we make the following assumptions: (a) the differences in projection data of adjacent angles are minor; (b) the noise in projection data of distinct projection angles is independent of each other and zero mean. Under those assumption, $\left|\left| F_{\theta}(y_s^k) - y_{s-1}^k\right|\right|_2^2 + \left|\left| F_{\theta}(y_s^k) - y_{s+1}^k\right|\right|_2^2$ is equal to $2*\left|\left| F_{\theta}(y_s^k) - x_{s}^k\right|\right|_2^2$. As a consequence, Eq. \ref{eq1} and Eq. \ref{eq2} are equivalent.

We can observe from Eq.\ref{eq1} that if we employ the adjacent two slices and itself as the supervision, the unsupervised problem is equal to having the ground truth as the supervision under specific assumptions.

Because our method is a training strategy rather than an architectures per se, it may be used to any acceptable neural network backbone. In this work, to implement our Noise2Projection, we use standard U-net \cite{ronneberger2015unet} with 32 basic feature-maps as the primary neural network, $F$, and Eq. \ref{eq1} as the loss function.  

\subsection{Datasets and Evaluation}
\begin{figure*}[htb]
	\begin{center}
		\includegraphics[scale=0.45]{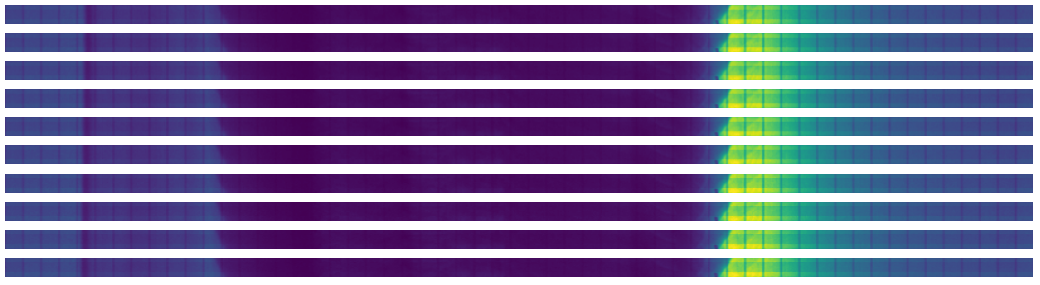}
		\caption{Example image of raw LDCT scan projection data from randomly selected 10 consecutive projection angles.}
		\label{120_30_qa1}  
	\end{center}
\end{figure*}

We trained our model by using clinical human chest CT scan datasets. A total of 10 clinic patient(range 44.3-103kg) CT scan data were taken by the Minfound ScintCare CT16 scanner, and we randomly selected 7, 1 and 2 patients data for training, validation and testing, respectively. The number of projection views for a full 360$^\circ$ rotation is 1024 and the dimension of each acquired projection image is 912$\times$16. On average, each patient has 17741 projections. During the projection data acquisition, the x-ray tube current was set at 30 mA, and the duration of the x-ray pulse at each projection view was 15 ms. The tube voltage was set to 120 kVp. The reconstruction was performed using the manufacturer-provided software with all physical corrections. The size of each reconstructed CT image is 512$\times$512 with pixel size of 0.706*0.706mm. Because a scan circle has 1024 projections, the difference between consecutive projection data is quite modest. In Fig. ~\ref{120_30_qa1}, 10 consecutive CT projections are shown, with no noticeable structure change. 

\subsection{Implementation Details}
In this work, we implement Noise2Projection and one unsupervised model (CycleWGANs \cite{zhou2020supervised}) in image domain  as a baseline. The unsupervised baseline is trained with the same loss function as in the original paper \cite{zhou2020supervised}, but without the supervised learning loss, just adversarial loss ($\mathcal{L}_{adv}$), consistency loss ($\mathcal{L}_{cyclic}$), and cyclic loss ($\mathcal{L}_{identity}$) are included.
 
The denoising network was built in Keras with Tensorflow backend and tested on NVIDIA GeForce RTX 3090 GPUs. The neural networks were initialized with a normal distribution and trained by Adam optimizer with with $\beta_1$ =0.5 and $\beta_2$=0.999 during 100 epochs. The learning rate was set at $2\times10^{-4}$ and decremented linearly to zero. For training, the batch size was set to 16 and for inference, it was set to 1. The network uses full size $912 \times 16$ projections image for end-to-end processing for training and inference.

\subsection{Evaluation Metrics}
No NDCT images can be utilized as a reference because our dataset was obtained from a clinical setting. In the CT image domain and projection domain, the noise reducation ratio, Z-profile, profile, and human vision sanity check were employed to assess model performance. 

\section{Experimental Results}

After the model has been trained, the raw projection data from the test set is fed into it our proposed model to produce the noise-reduced projection data. The quantitative characteristics of the projected data and the reconstructed CT images were examined in the following investigation.

\subsection{Experimental Results on Projection Data}
\begin{figure*}[htb]
	\begin{center}

	\subfigure[Raw and Noise2Projection model processed projection image]{	
	\includegraphics[scale=.535]{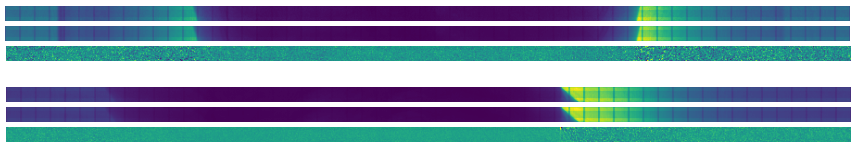}
	}
	\subfigure[Raw Projection Image]{
	\includegraphics[scale=.24]{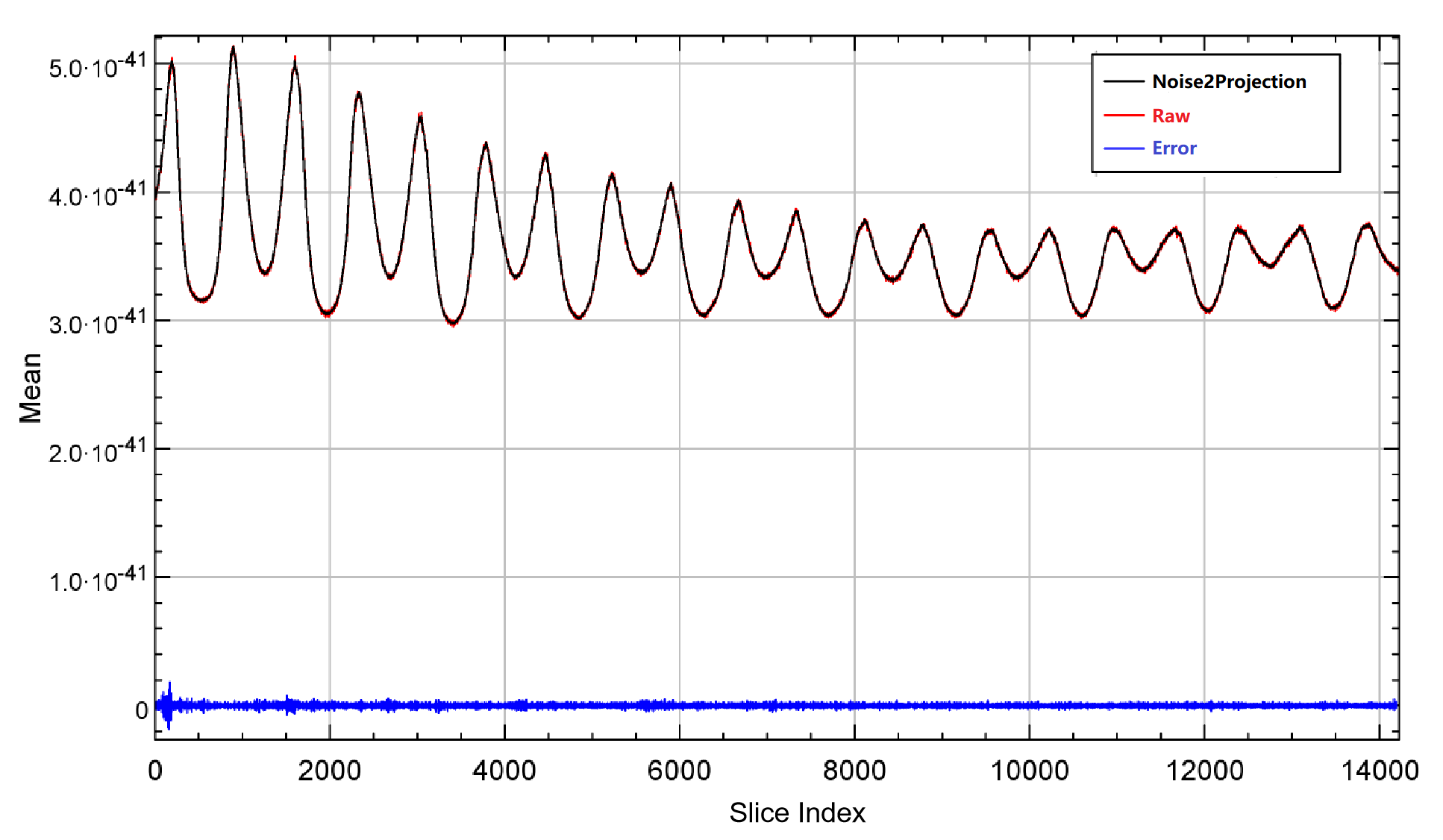} 
}
	\caption{Noise2Projection model processed results. (a) Two sample of raw and Noise2Projection model processed projection image with corresponding error (Raw, Noise2Projection, and the difference between them are listed from top to bottom). (b) The Z-profile for Raw and Noise2Projection processed projection image and corresponding error. }
	\label{120_30_cm}  
	\end{center}
\end{figure*}

Fig. \ref{120_30_cm}(a) shows a comparison of the raw and Noise2Projection processed projection images. Because the projected image's height is only 16 pixels, it's tough to see the details, but the difference between the two may be seen. The Noise2Projection model predicts a decreased noise level in the projection data and no change in the local structure, as shown in Fig. \ref{120_30_cm}(a).

The Z-profile (The mean value of each slice along the Z-direction) of raw and Noise2Projection processed projection data are shown in Fig.~\ref{120_30_cm}(b). Due to low-dose CT scan, the number of photon passing through the body is lower than NDCT scan and resulting noise projection data. The noise in Noise2Projection model estimated data is lower than the raw projection data (smaller fluctuation in Z-profile), and the quantitative inaccuracy is almost non-existent (relative error is less than 0.01\% )

\subsection{Experimental Results on CT Image}
Our Noise2Projection model first processes the raw projection data, then we import the processed projection data along with the raw projection data into the CT vendor's reconstruction tool for image reconstruction, and finally we assess the quality of the reconstructed images using quantitative and qualitative methods.

\subsubsection{Artifact removal}
\begin{figure*}[htb]
	\begin{center}
		\subfigure[LDCT]{
			\includegraphics[scale=.42]{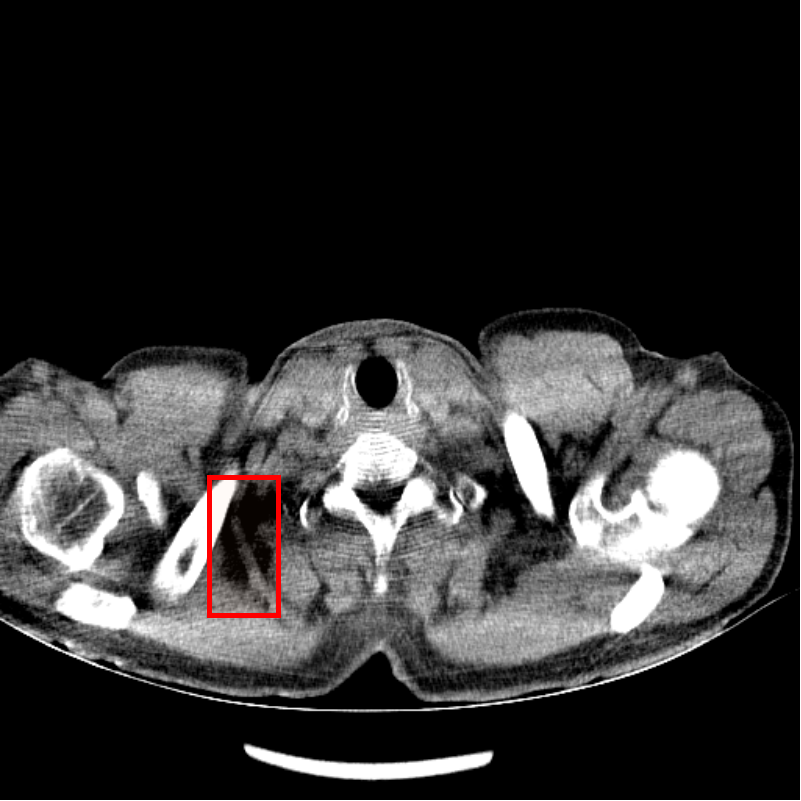}
		}
		\subfigure[Noise2Projection]{	
			\includegraphics[scale=.42]{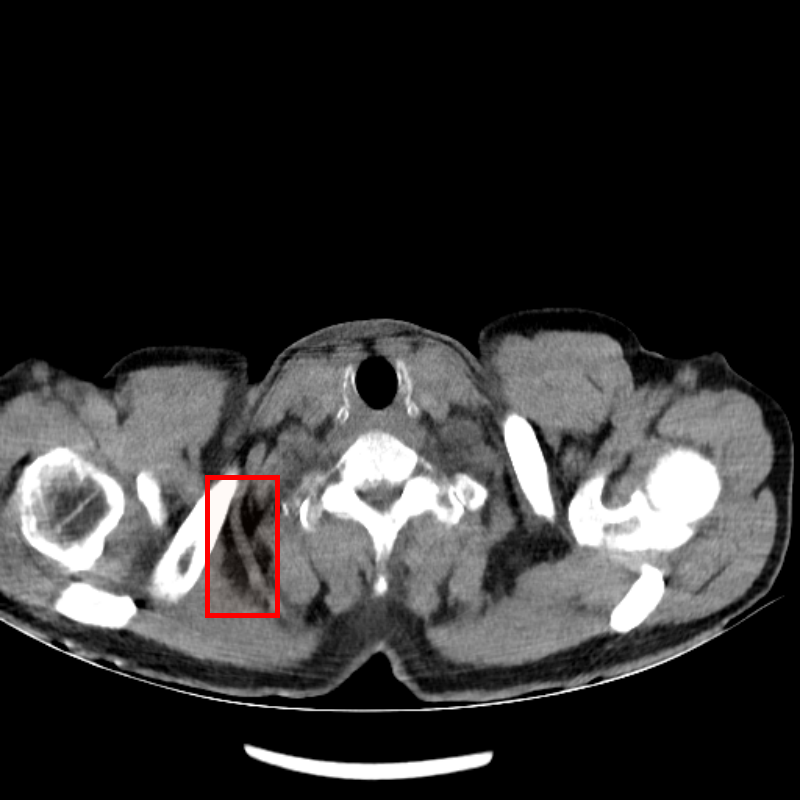}
		}
		\subfigure[CycleWGANs]{	
			\includegraphics[scale=.42]{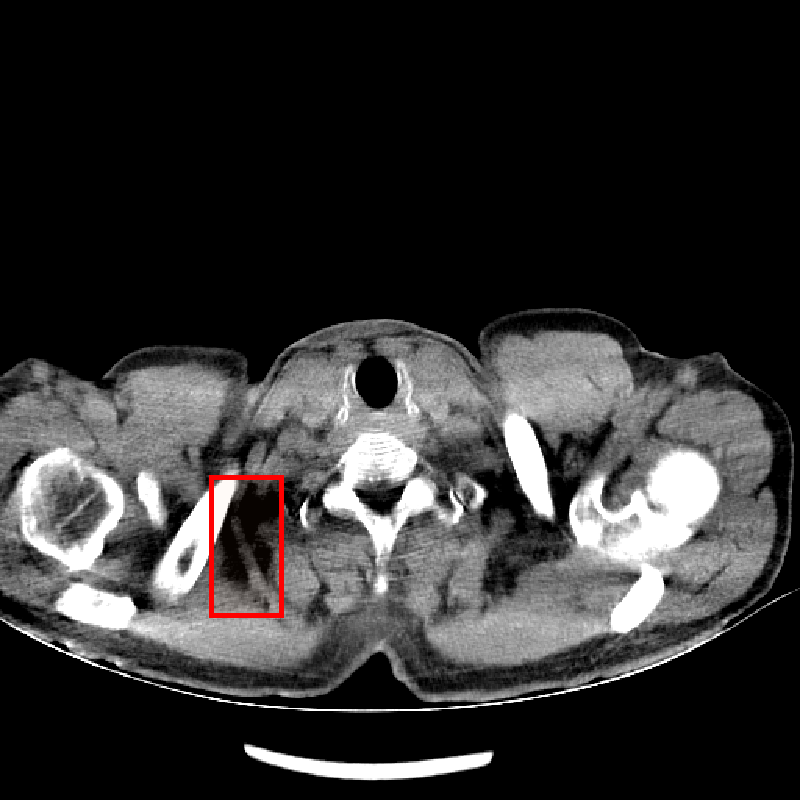}
		}	
		\subfigure[NLM]{	
			\includegraphics[scale=.42]{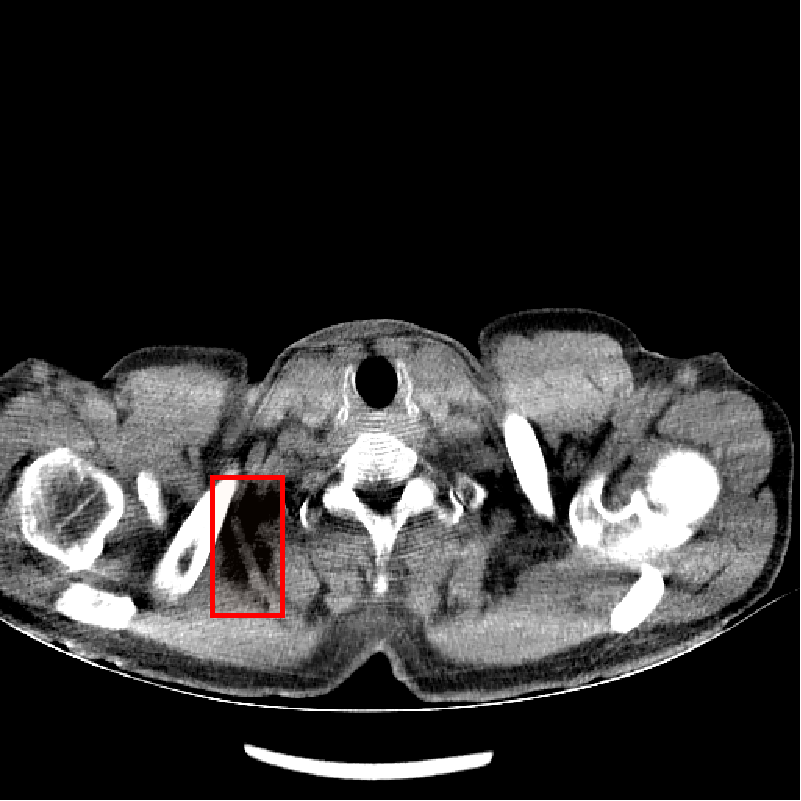}
		}	
		\caption{Selected CT images from: (a) LDCT, (b) Noise2Projection, (c) CycleWGANs, (e) NLM. WL/WW is 40/320}
		\label{120_30_test}  
	\end{center}
\end{figure*}

Fig. \ref{120_30_test} shows CT images reconstructed from raw LDCT projection data and projection data with Noise2Projection model processing. The LDCT image is shown in Fig.\ref{120_30_test} (a) and the Noise2Projection result is shown in Fig.\ref{120_30_test} (b). There are some artifacts, including "black bands" and ring artifacts, can be seen in the original LDCT image, which are likely to obstruct the physician's diagnosis. This is because the x-ray flux is reduced, resulting in fewer photons being detected. These artifacts are most common in the shoulder and neck region due to the greater bone content in this location, which results in stronger photon attenuation.
As seen in Fig. \ref{120_30_test}, our model processing efficiently removed the "black bands" and ring artefacts in the LDCT images (mainly from bone artifacts, as indicated by red rectangles in Fig. \ref{120_30_test}(a) and (b)) while also recovering part of the destroyed structures.

\subsubsection{Noise reduction}
We selected ROIs in uniform locations on numerous organs or tissues on the reconstructed images and evaluated the mean CT values and standard deviation within each ROI region to assess the efficacy of our proposed approach on noise reduction in CT images. Table \ref{tab1} illustrates the mean CT values and standard deviation of ROIs in numerous organs, as well as the noise reduction ratios. 
We can observe that our proposed strategy is quite effective in the liver, muscle, and kidney with noise reduction ratios of 47.68\%, 44.58\%, and 44.63\%, respectively. The lungs, on the other hand, do not show considerable noise reduction since they are mostly air, with minimal x-ray attenuation.

\begin{figure*}[htb]
	\begin{center}
		\subfigure[LDCT image]{
			\includegraphics[scale=.21]{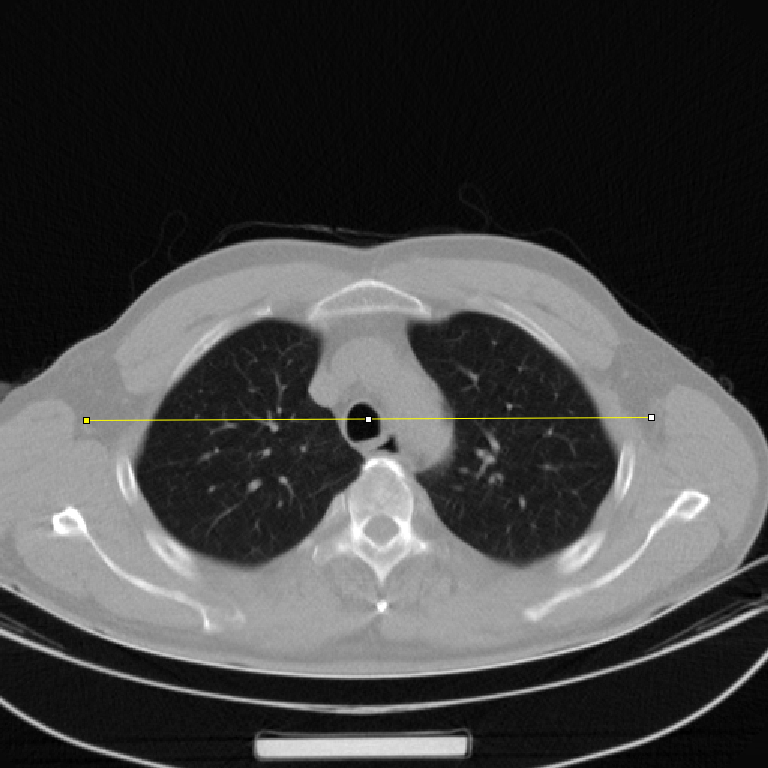}
		}
		\subfigure[Noise2Projection processed LDCT image]{	
			\includegraphics[scale=.10]{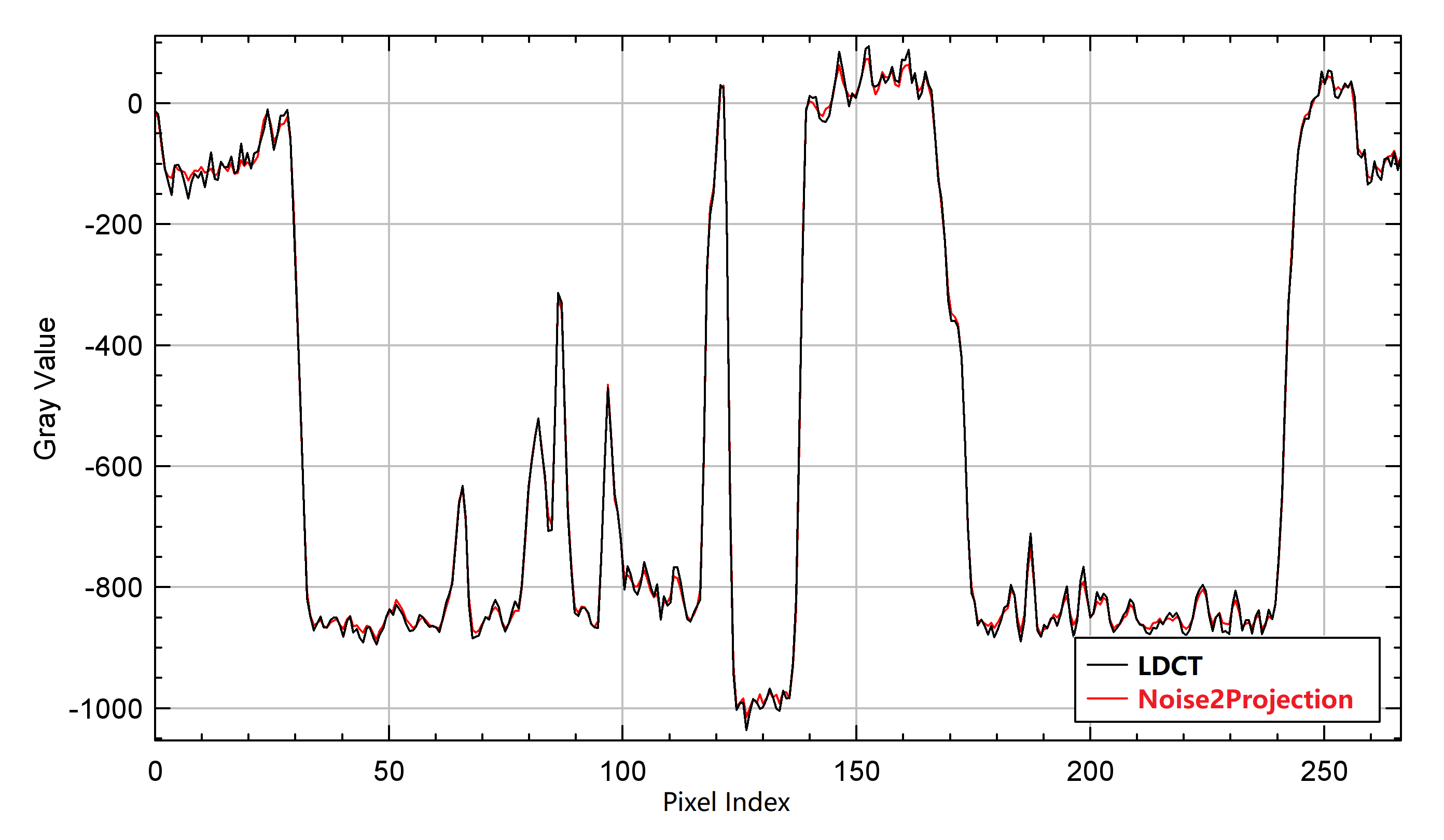}
		}
		
		\caption{The profiles through the lung in LDCT and Noise2Projection images. (a) LDCT image and position of the profile line, (b) comparison of line profile in LDCT and Noise2Projection processed CT images.}
		\label{120_30_profile}  
	\end{center}
\end{figure*}

In addition, We also looked at the changes in line-profiles on the reconstructed images. We estimated the line-profile from a random selected image with no noticeable artifacts. Fig. ~\ref{120_30_profile} displays a line-profile comparison. We can see that the line-profile generated by the Noise2Projection model matches the LDCT extremely well, and that the fluctuation is decreased (means smaller noise).

\begin{table}
	\caption{Mean and standard deviation of CT value in each organ and corresponding noise reduction ratio(NRR). }\label{tab1}
	\begin{tabular}{|l|l|l|l|l|}
		\hline
		Image Name &  Lung & Liver & Muscle & Kidney \\
		\hline
		LDCT &  -832.83 $\pm$ 80.74 & 60.89 $\pm$ 15.54 & 59.42 $\pm$ 20.32 & 33.74 $\pm$ 14.92\\
		CycleWGANs & -830.86 $\pm$ 79.28 & 61.10 $\pm$ 12.23& 56.92 $\pm$ 14.79& 36.50 $\pm$ 10.57\\
		NLM & -832.87 $\pm$ 80.15 & 62.38 $\pm$ 14.04& 59.38 $\pm$ 16.69& 33.81 $\pm$ 11.94\\
		Noise2Projection &  -833.09 $\pm$ 77.01 & 58.51 $\pm$ 8.13 & 58.69 $\pm$ 11.26 & 29.41 $\pm$ 8.26\\
		\hline
		NRR for CycleWGANs& 1.81\% & 21.30\% & 27.21\% & 29.15\%\\
		NRR for NLM& 0.73\% & 9.65\% & 17.86\% & 19.97\%\\
		NRR for Noise2Projection& \bfseries4.62\% & \bfseries47.68\% & \bfseries44.58\% & \bfseries44.63\%\\
		\hline
	\end{tabular}
\end{table}

\subsubsection{Comparison with image domain methods}
To study the effectiveness of our proposed model, we compared it with one deep learning mode, CycleWGANs\cite{zhou2020supervised}, and one traditional methods, Non-local Mean(NLM)\cite{dutta2013non}. Both of the two methods mentioned above are high-performance methods in the image domain.
Raw LDCT, Noise2Projection, CycleWGANs, and NLM processed CT images are compared in Fig. \ref{120_30_test} and Table \ref{tab1}. Both CycleWGANs and NLM have strong noise reduction effects, with noise reduction ratios ranging from 1.81\% to 29.15\% for CycleWGANs and 0.73\% to 19.97\% for NLM, respectively. We should also mention that neither CycleWGANs nor NLM are capable of properly removing the artifacts found in LDCT images, such as "black bands" and ring artifacts. Our Noise2Projection model not only has the best noise reduction performance, but it also effectively removes artifacts from LDCT images, demonstrating promising results.

\section{Conclusion}
In this research, we propose to employing a self-supervised learning model called Noise2Projection to reconstruct low-dose CT images from raw projection data. We train and evaluate the model using clinical data, and quantitative and qualitative results suggest that our method outperforms two image domain methos. The reconstructed CT images are not only substantially less noisy, but also have less artifacts. The suggested research may lead to the development of a new set of LDCT image denoising methods. In our future work, we will investigate the mechanism of artifact reduction that was observed in this research, and we will apply our method to other artifact reduction tasks in CT and PET scans.

\subsubsection{Acknowledgements} ******

%
%
%
\bibliographystyle{splncs04}
\bibliography{./myreference2}

\begin{thebibliography}{10}
\providecommand{\url}[1]{\texttt{#1}}
\providecommand{\urlprefix}{URL }
\providecommand{\doi}[1]{https://doi.org/#1}

\bibitem{2012Image}
Image quality and radiation dose of low dose coronary ct angiography in obese
  patients: Sinogram affirmed iterative reconstruction versus filtered back
  projection. European Journal of Radiology  \textbf{81}(11) (2012)

\bibitem{2019Noise2Self}
Batson, J., Royer, L.: Noise2self: Blind denoising by self-supervision  (2019)

\bibitem{2012Iterative}
Beister, M., Kolditz, D., Kalender, W.A.: Iterative reconstruction methods in
  x-ray ct. Physica Medica  \textbf{28}(2),  94--108 (2012)

\bibitem{dutta2013non}
Dutta, J., Leahy, R.M., Li, Q.: Non-local means denoising of dynamic pet
  images. PloS one  \textbf{8}(12),  e81390 (2013)

\bibitem{gnudi2020denoising}
Gnudi, P., Schweizer, B., Kachelrie{\ss}, M., Berker, Y.: Denoising of x-ray
  projections and computed tomography images using convolutional neural
  networks without clean data. In: The 6th International Conference on Image
  Formation in X-Ray Computed Tomography. pp. 590--593 (2020)

\bibitem{2009Iterative}
Hara, A.K., Paden, R.G., Silva, A.C., Kujak, J.L., Lawder, H.J., Pavlicek, W.:
  Iterative reconstruction technique for reducing body radiation dose at ct:
  feasibility study. Ajr Am J Roentgenol  \textbf{193}(3),  764--771 (2009)

\bibitem{Hu2017Low}
Hu, Chen, Yi, Zhang, Weihua, Zhang, Peixi, Liao, Ke, Li: Low-dose ct via
  convolutional neural network. Biomedical Optics Express  (2017)

\bibitem{2009Computed}
Jiang, H.: Computed Tomography Principles, Design, Artifacts, and Recent
  Advances, 2nd Edition. Computed Tomography: Principles, Design, Artifacts,
  and Recent Advances, Second Edition (2009)

\bibitem{2005Sinogram}
Jing, W., Lu, H., Li, T., Liang, Z.: Sinogram noise reduction for low-dose ct
  by statistics-based nonlinear filters. International Society for Optics and
  Photonics  (2005)

\bibitem{2004Strategies}
Kalra, M.K., Maher, M.M., Toth, T.L., Hamberg, L.M., Blake, M.A., Jo-Anne, S.,
  Sanjay, S.: Strategies for ct radiation dose optimization. Radiology
  \textbf{230}(3),  619--628 (2004)

\bibitem{2013Image}
Kang, D., Slomka, P., Nakazato, R., Woo, J., Berman, D.S., Kuo, C., Dey, D.:
  Image denoising of low-radiation dose coronary ct angiography by an adaptive
  block-matching 3d algorithm. In: Spie Medical Imaging (2013)

\bibitem{2019Noise2Void}
Krull, A., Buchholz, T.O., Jug, F.: Noise2void - learning denoising from single
  noisy images. In: 2019 IEEE/CVF Conference on Computer Vision and Pattern
  Recognition (CVPR) (2019)

\bibitem{2019High}
Laine, S., Karras, T., Lehtinen, J., Aila, T.: High-quality self-supervised
  deep image denoising  (2019)

\bibitem{2018Noise2Noise}
Lehtinen, J., Munkberg, J., Hasselgren, J., Laine, S., Karras, T., Aittala, M.,
  Aila, T.: Noise2noise: Learning image restoration without clean data  (2018)

\bibitem{2009Projection}
Manduca, A., Yu, L., Trzasko, J.D., Khaylova, N., Kofler, J.M., Mccollough,
  C.M., Fletcher, J.G.: Projection space denoising with bilateral filtering and
  ct noise modeling for dose reduction in ct. Medical Physics  \textbf{36}(11),
   4911--4919 (2009)

\bibitem{papkov2021noise2stack}
Papkov, M., Roberts, K., Madissoon, L.A., Shilts, J., Bayraktar, O., Fishman,
  D., Palo, K., Parts, L.: Noise2stack: Improving image restoration by learning
  from volumetric data. In: International Workshop on Machine Learning for
  Medical Image Reconstruction. pp. 99--108. Springer (2021)

\bibitem{ronneberger2015unet}
Ronneberger, O., Fischer, P., Brox, T.: U-net: Convolutional networks for
  biomedical image segmentation (2015)

\bibitem{20183D}
Shan, H., Zhang, Y., Yang, Q., Kruger, U., Kalra, M.K., Sun, L., Cong, W.,
  Wang, G.: 3d convolutional encoder-decoder network for low-dose ct via
  transfer learning from a 2d trained network. IEEE Transactions on Medical
  Imaging  \textbf{37}(6), ~1522 (2018)

\bibitem{1987Noise}
Snyder, D.L., Miller, M.I., Thomas, L.J., Politte, D.G.: Noise and edge
  artifacts in maximum-likelihood reconstructions for emission tomography. IEEE
  Transactions on Medical Imaging  \textbf{6}(3), ~228 (1987)

\bibitem{2018Low}
Yang, Q., Yan, P., Zhang, Y., Yu, H., Shi, Y., Mou, X., Kalra, M.K., Zhang, Y.,
  Sun, L., Wang, G.: Low-dose ct image denoising using a generative adversarial
  network with wasserstein distance and perceptual loss. IEEE Transactions on
  Medical Imaging pp. 1348--1357 (2018)

\bibitem{zainulina2021no}
Zainulina, E., Chernyavskiy, A., Dylov, D.V.: No-reference denoising of
  low-dose ct projections. In: 2021 IEEE 18th International Symposium on
  Biomedical Imaging (ISBI). pp. 77--81. IEEE (2021)

\bibitem{zhang2021noise2context}
Zhang, Z., Liang, X., Zhao, W., Xing, L.: Noise2context: Context-assisted
  learning 3d thin-layer for low-dose ct. Medical Physics  \textbf{48}(10),
  5794--5803 (2021)

\bibitem{zhou2020supervised}
Zhou, L., Schaefferkoetter, J.D., Tham, I.W., Huang, G., Yan, J.: Supervised
  learning with cyclegan for low-dose fdg pet image denoising. Medical image
  analysis  \textbf{65},  101770 (2020)

\end{thebibliography}

\end{document}